\documentclass[iop]{emulateapj}
\usepackage{natbib}
\newcommand{\Kepler}{{\sl Kepler}\ }

\newcommand{\be}{\begin{equation}}
\newcommand{\ee}{\end{equation}}

\slugcomment{Reviewed by ApJ, Awaiting Re-submission}
\shorttitle{ Circumbinary Instability }
\shortauthors{Sutherland and Fabrycky}

\begin{document}

\title{
On the Fate of Unstable Circumbinary Planets: \\
Tatooine's Close Encounters with a Death Star
}
\author{
Adam P. Sutherland\altaffilmark{1},
Daniel C. Fabrycky\altaffilmark{1}
}
\altaffiltext{1}{Department of Astronomy and Astrophysics, University of Chicago, 5640 S. Ellis Ave., Chicago, IL 60637, USA}
\email{adamsutherland@uchicago.edu; fabrycky@uchicago.edu}

\begin{abstract}
Circumbinary planets whose orbits become unstable may be ejected, accreted, or even captured by one of the stars.  We quantify the relative rates of these channels, for a binary of secondary star's mass fraction $0.1$ with an orbit of $1$~AU.  The most common outcome is ejection, which happens $\sim 80\%$ of the time.  If binary systems form circumbinary planets readily and sloppily, this process may fill the Milky Way with free-floating planets.  A significant fraction of the time, $\sim 20\%$, the unstable planet strikes the primary or secondary.  We tracked whether a Jupiter-like planet would undergo tidal stripping events during close passages, and find that these events are not strong enough to change the trajectory of the planet, though this may be observable from a changed structured for free-floating planets that are produced by this process. 

\end{abstract}


\section{Introduction} 
\label{secIntro}

The recent unambiguous detections of circumbinary planets \citep{2011Doyle, 2014Welsh} have given fresh inspiration for theories of planet formation and evolution in such a perturbed environment. Most theorists have concluded that planet formation could not have occurred within close proximity to the binary star. A number of models suggest that in-situ formation of the circumbinary \Kepler planets is highly unlikely and that formation outside followed by inward migration is the most probable formation theory \citep{2007Scholl,Meschiari2012,Rafikov2013,Silsbee2015}.  Although nearby the streamlines of a circumbinary disk may be altered to a steady-state which is amenable to planet formation, the chaotic dynamics near the binary could deplete the disk to a level that is not amenable to planet formation \citep{2015BromleyKenyon}.

If indeed gas giants are formed at several AU within a circumbinary disk, they must undergo migration to attain their observed orbits, within about 1 AU.  An early prediction of such theories is that planets may pile up near resonances, just outside the unstable region \citep{2003Nelson}. Moreover, Saturn-mass planets may survive just outside of resonance, whereas Jupiter-mass and larger planets might migrate into the unstable region \citep{2007Pierens, 2008Pierens}. In fact, the observations are consistent with both a pile-up just outside the unstable region and a preference for Saturn-mass planets. All of the \Kepler circumbinary planets have a mass or radius smaller than Jupiter and are located 1.09 and 1.46 times the critical instability radius \citep{2014Welsh}. Despite a low number of discovered circumbinary systems, this pile-up is thought to be physical and not a remnant of discovery bias \citep{Martin2014}. The physical pile up together with the difficulty of in situ formation suggests that migrating planets come to rest near the instability zones or that those who migrate farther are eliminated.  It also raises a host of questions about planet evolution in circumbinary systems \citep{2014Armstrong, Martin2014}.

The unstable systems studied here would also contribute to the recent observation of free-floating planets \cite{2011Sumi}. Gravitational microlensing revealed a number of unbound planets with an unknown formation process. In section \ref{sec:ffp} we discuss unstable circumbinary planet ejection and its ability create free floating planets. We also discuss its dependence on migration rate and formation as well as potential visible signatures of ejected circumbinary planets.

In this paper, we explore what happens to circumbinary planets that go unstable. An early survey of parameter space was provided by \cite{1999HW} – hereafter HW. We reproduce some results of HW in section~\ref{sec:traj}, also reporting the number of close encounters of the planet and the stars final fate of the planet. In section~\ref{sec:capture}, we add a tidal prescription and found that the individual stars may capture the planet into a close orbit. Section~\ref{sec:disc} draws some conclusions from these results, discussing observational signatures of unstable circumbinary planets. 

\section{Trajectories}
\label{sec:traj}

\subsection{Method}

In order to explore the location of stable orbits, we set up nu
ical integrations using a Bulirsch-Stoer algorithm ({\it Mercury}; \citealt{1999Chambers}) with various binary systems. We restrict the system to a planet on a coplanar orbit with respect to a binary of semimajor axis $a=1$~AU, a primary of mass $M_1=1$~$M_\odot$, and a secondary mass fraction of $\mu = M_2/(M_1+M_2)=0.1$. We varied the eccentricity between $0.0$ and $0.3$ with $0.1$ intervals. The binary is initially at periastron with $\varpi=0^\circ$.  The planets are modeled as massless test particles positioned on a grid of distances, $1$~AU to $4$~AU with spacing $0.005$~AU, from the barycenter of the binary.  The initial longitude was chosen in equal numbers between $0^\circ$, $90^\circ$, $180^\circ$, and $270^\circ$ . The planets initially have a circular orbit around the barycenter (i.e., $e=0$ in Jacobian elements) and the integration lasts $10^4$ years. This timescale, which is slightly greater than the $10^4$ binary periods in HW, proved to be sufficient time to cause most unstable orbits to be removed from the system, by ejection or collision (although see section~\ref{sec:instab}b and figure~\ref{fig:lifetime}, below). We specify a close encounter as any approach that passes within $0.05$~AU from a star at its closest point (approximately where tides begin to modify orbits). {\it Mercury} is designed for central body systems and does not record close encounters with the central body. A workaround that allows Mercury to report close encounters with both stars is to essentially remove the central body from the system and to specify both stars within the {\it big.in} file. {\it Mercury} does not allow the central body to be massless so we placed an incredibly small and nearly massless ``central body'' on a circular orbit $5$~AU away from the binary.  With this approach, an unpleasant side-effect is caused by {\it Mercury}'s scaling of the physical radius of bodies by the Hill radius of the central body.  So, in order to avoid point-particle stars, the physical radius of each star, 0.004649 and 0.0004649 ~AU, was hardwired into the MCE\_INIT.FOR subroutine. Any planet that passes more than $100$~AU from the central body is removed from the integration and marked as ejected. 

\subsection{Results}

\begin{figure*}[htp]

  \centering

  \begin{tabular}{cc}


    \includegraphics[width=85mm]{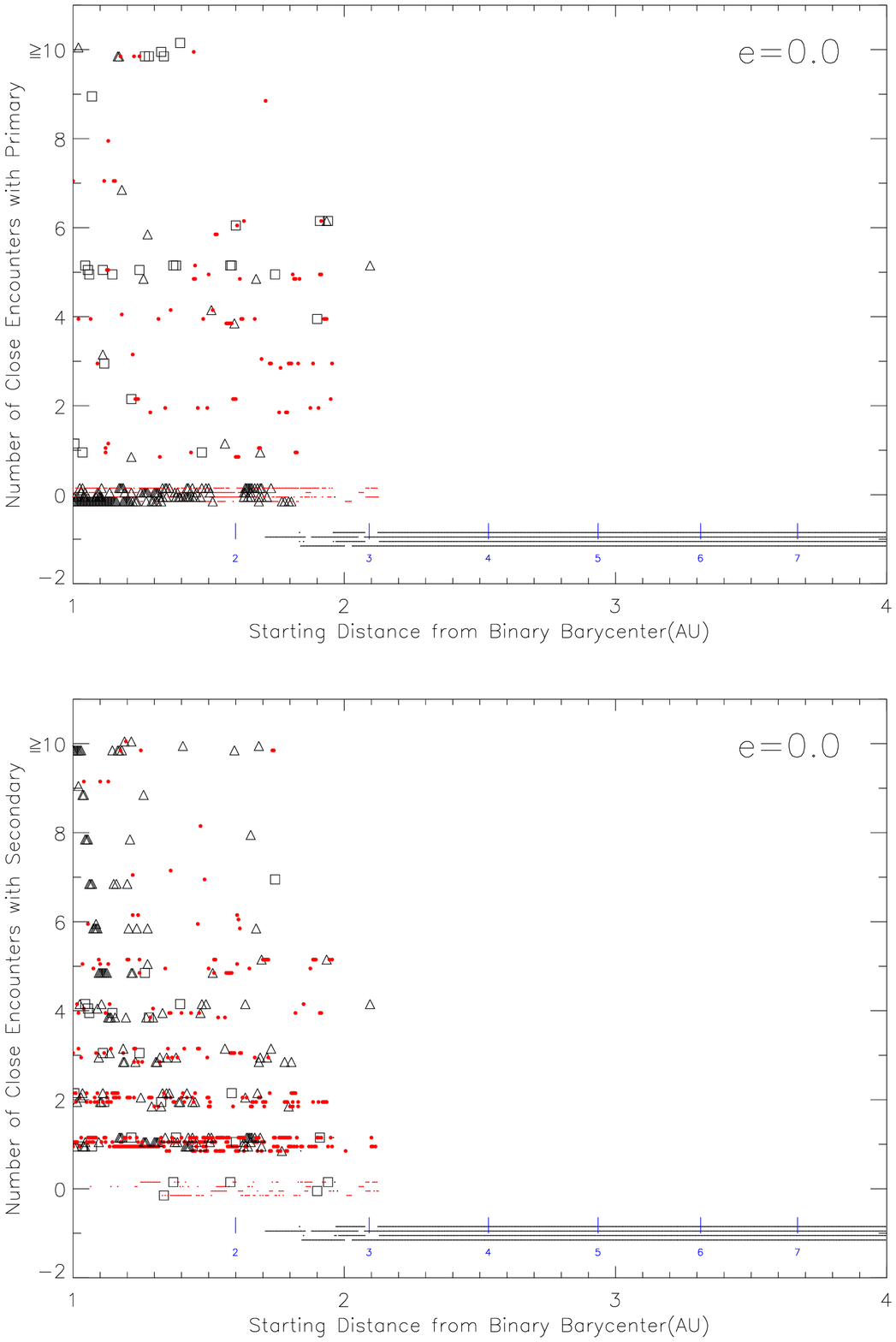}&

    \includegraphics[width=85mm]{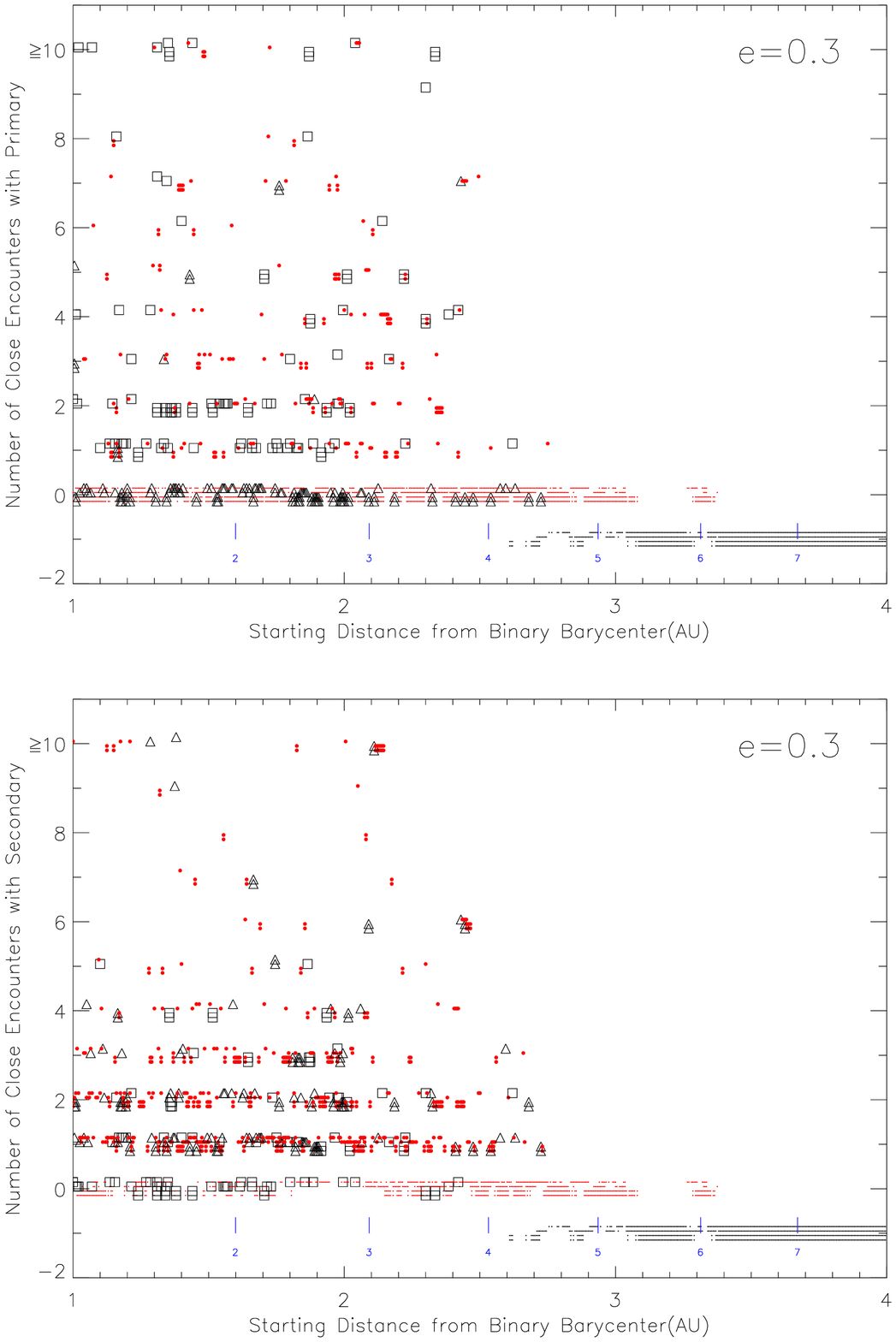}

  \end{tabular}

\caption{ Number of close encounters with the primary {\it top} and secondary {\it bottom} versus initial distance from binary.   {\it Left:} binary $e=0.0$; {\it Right:} binary $e=0.3$.  The horizontal axis is the starting position of the planet relative to the barycenter of the binary. The vertical axis is the number of close encounters each planet had.  All planets that had more than $10$ close encounters appear as having 10 close encounters to increase the readability of the graph. The black dots are the planets that survived the full $10^4$ years integration; those survivors that had zero close encounters are offset by $-1$ in order to better distinguish them from the ejections. Ejections are marked in red and all ejections with at least one close encounter are represented with a large dot to increase readability. Planets that impacted the primary are represented by a square and those that impacted the secondary are represented by a triangle. The plot contains the results of all $4$ starting longitudes each offset by $0.1$ centered on the number of close encounters, ordered $0^\circ$, $90^\circ$, $180^\circ$, and $270^\circ$, top to bottom. Integer period resonances are labeled in blue.   \label{fig:e0} \vspace{0.2 in} }
\end{figure*}

The fates of the planets are given in figure~\ref{fig:e0}, for eccentricities of $0.0$ and $0.3$.  These cases show the trends graphically, but we also include $e=0.1$ and $0.2$ in the numerical results below. 

\subsubsection{Instability}
\label{sec:instab}
{\it a. Critical Semimajor axis}

There exists a minimum distance at which planets survive the full time integration from all starting longitudes, referred to as the critical semimajor axis by HW. This is not the semimajor axis beyond which all orbits are stable, since islands of instability exist outside of the critical orbit. These islands of instability are located at the integer resonances. As eccentricity increases, the critical semimajor axis increases and the orbits between the island and the critical semimajor axis become unstable, and the next integer resonance becomes unstable.  For the integrations of $10^4$ years, most systems had only one island of instability outside the critical semimajor axis. When determining the stability of an orbit of a planet around binary stars of a certain mass ratio, the two main factors to consider are time and eccentricity. As eccentricity increases, so does instability.  Most planets were removed in the first $1000$~years, but some did take longer to become unstable. Integrating longer would prove to find more unstable orbits but the would not likely change the results significantly even if the integration lasted an order of magnitude longer.

{\it  b. Comparison to Previous Work}

\begin{figure}
\plotone{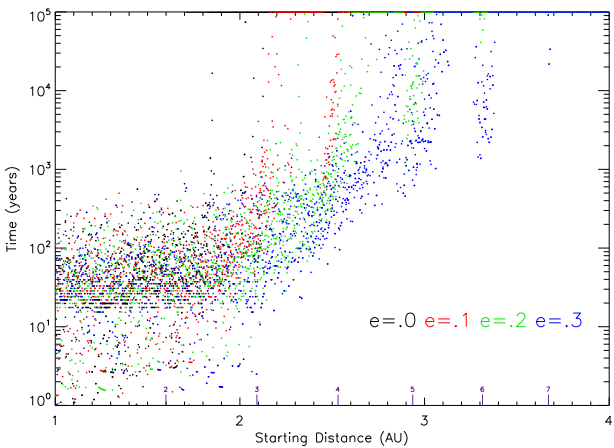} 
\caption{  
Ejection or collision time versus starting distance. Starting distance is the distance from the binary barycenter in AU.  Eccentricity $= 0.0$ in black, $0.1$ in red, $0.2$ in green, and $0.3$ in blue. Integer period resonances are marked on the horizontal axis.
\vspace{0.0 in} }
\label{fig:lifetime}
\end{figure}

The results differ from HW in a few ways. The integrations run slightly longer than those in HW. Our critical semimajor axis is somewhat closer than theirs due to our increased density of starting planets and the fact that we used four different longitudes rather than eight.  HW only presented $10^4$ years results. When the runtime was increased to $10^5$ years, slightly more planets were ejected or had collisions, see figure~\ref{fig:lifetime}. The increased time slightly increased the critical semimajor axis, removed surviving orbits inside of the critical semimajor axis, enlarged the island of instability, and even resulted in the ejection of planets at further resonances.\footnote{Although even after only $10^4$~yr, for eccentricity of $0.3$, one single planet near the 7th resonance was ejected and there is an island of instability on the 6th resonance; see figure~\ref{fig:e0}.}  The instability due to resonance takes longer to remove a planet from the system because the initially circular orbit of the planet becomes perturbed very slowly before finally becoming unstable enough to be ejected, compared with the planets that start close to the binary what are made unstable very quickly because of their proximity to the stars.  

\subsubsection{Final Fate}

\begin{figure}
\plotone{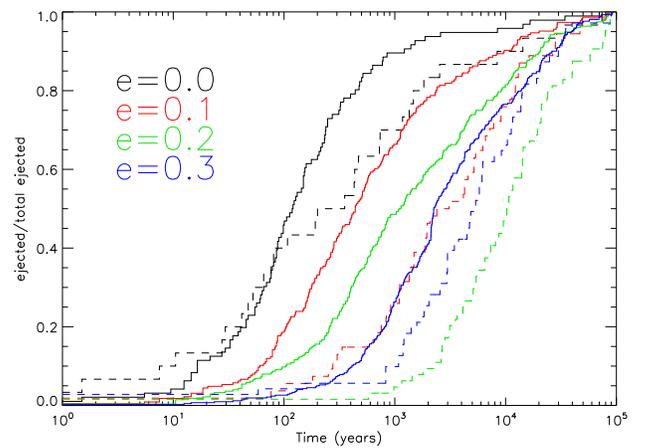} 
\caption{  
Fraction of total ejected planets for islands of instability, dashed, and all ejections more distant than the most distant impact outside the island of instability, solid, versus time. Eccentricity $= 0.0$ in black, $0.1$ in red, $0.2$ in green, and $0.3$ in blue.
\vspace{0.0 in} }
\label{fig:ejecttime}
\end{figure}

Ejections outnumber collision in every setup: $81.0\% \pm 0.6\%$ of all unstable orbits result in ejection. Collisions with the secondary are more common than collisions with the primary: $68.2\% \pm 1.6\%$ of collisions are with the secondary. For the starting distance we consider, collisions with the primary become more common when eccentricity increases ($14.9\% \pm 2.7\%$ of collisions for $e=0$ and $42.2\% \pm 3.1\%$ of collisions for $e=0.3$).  Planets that have three or more close encounters with a body are much more likely to impact that same body. Very few (only 7, all in the $e=0$ run) had a close encounter during the $10^4$ year integration and yet survived. The time for ejection is listed for both the islands of instability and the all ejected planets after the most distance impact is shown in \ref{fig:ejecttime}.

\epsscale{1.20}
\epsscale{1.0}

\epsscale{1.20}
\epsscale{1.0}

\subsubsection{Number of Close Encounters}

Most planets had fewer than $10$ close encounters but several had over $150$. Except for the zero-eccentricity case, all of the ejections in the first resonant island of instability result without any close encounters with the binary. This is most likely because the island of instability outside the critical semimajor axis is much closer to the binary at an eccentricity of zero than the islands of instability in the systems with some eccentricity.
\vspace{0.4 in}

\section{Tidal Disruption and Capture}
\label{sec:capture}

\subsection{Method}

During {\it Mercury}'s integrations, the only force is $N$-body gravitation. In reality, during all close encounters, tides act on the planet, altering its orbit. 

We first tried adopting the equilibrium tidal dissipation given in \cite{2012BN}, their equation~9, by adding a user-defined force in the MFO\_USER.FOR subroutine. The tidal dissipation force altered the results slightly, but did not increase the number of close encounters or capture any planets.  Due to the violent nature of the unstable close encounters, the planets did not stay on close-passing orbits long enough for dynamical or equilibrium tides to alter the orbits much, certainly not enough to result in a capture. 

\begin{figure}
\plotone{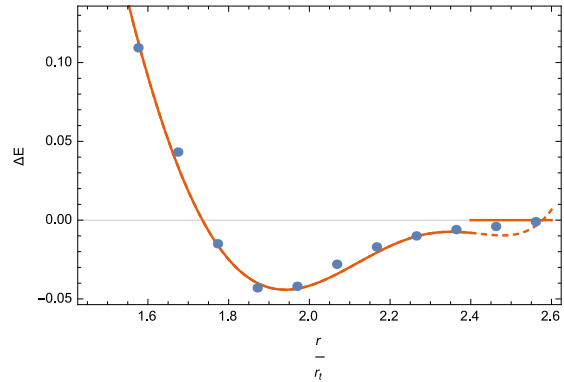} 
\caption{  
Change in energy due to tidal interaction and distance from the star at closest approach. The horizontal axis is the distance in $\boldmath{\frac{r}{r_t}}$ and the vertical axis is the energy in $3.37*10^{43}$ erg or $3.61*10^{-4}$ $M_{Jup}^2$ $R_{Jup}^2$ / $day^2$.
\vspace{0.2 in} }
\label{fig:poly}
\end{figure}

When a planet passes very close to a star, within its Roche limit at periastron, on a parabolic or nearly parabolic orbit, mass can be stripped from the planet and the orbital energy can be altered so that one of the stars could potentially capture the planet. Adopting the changes in orbital energy presented in \cite{2005Faber}, we edit {\it Mercury} to account for these very close approaches. At each close encounter that passed within a few characteristic tidal radii (defined following \citealt{2013Liu}),
\begin{equation}
r_t = (M_\star/M_p)^{1/3} R_p,
\end{equation}
where $R_p$ is the planet radius and $M_\star$ and $M_p$ are the stellar and planetary masses, the velocity of the planet was altered according to the energy data given in \cite{2005Faber}.  This was accomplished by editing the MCE\_STAT.FOR subroutine so during every close encounter the planet received a kick along its velocity vector to correspond to the specific energy change due to the close encounter. We fit the change-in-energy results from \cite{2005Faber}, their table~1, to a polynomial for a planet with Jupiter's characteristics --- see figure~\ref{fig:poly}.  We assume that the energy kick depends only on the closest approach distance $r_p$ in units of $r_t$, and not, for instance, the particular eccentricity of the initial orbit, which determines the specific shape of the orbit near closest approach where the stripping takes place. 
This curve was reproduced by \cite{2013Liu}, who studied planets with cores, so we believe it is a robust prediction that planets can become more tightly bound when they are stripped. The results were qualitatively verified by the additional method of grid hydrodynamical codes in \cite{2011Guillochon} rather than smoothed-particle hydrodynamical codes in \cite{2005Faber} and \cite{2013Liu}. We apply this change in energy when the the trajectory's closest approach distance is less than $2.4 r_t$, or $0.01127$~AU from the primary.  Planets that pass closer than about $1.6 r_t$ may have a tremendous increase in specific orbital energy, leading to their ejection, according to \cite{2005Faber}.  On the other hand, \cite{2013Liu} found that planets with cores may increase specific orbital energy when they pass within $1.6 r_t$, though avoid ejection, since the inclusion of a core resists the planet's stripping.  Therefore, for all orbits that pass very close to the star calling for a very large velocity change, we take a middle path, setting the planet's velocity equal to the escape velocity with respect to that star in the event that the prescribed velocity is greater than the escape velocity.  (It may either remain bound to the binary system, or be ejected, depending on how it interacts with the other star.)    Thus we alter the velocity according to:
\begin{equation}
v_{new} = {\rm min}[ (v_{old}^2+2 \Delta KE / m_p)^{1/2} , (2 G M_\star / r_p)^{1/2} ]
\end{equation}
for approaches closer than $2.4 r_t$.   

If during these tidal encounters closer than $2.4r_t$, the planet becomes progressively more bound, we mark it as captured and removed from the integration when it attains an orbit of less than $a_p=0.05$~AU around the single star. At this point, the planet is deep within the gravitational influence of the star that has captured it, and it cannot be scattered away by the other star. Also, in this modified code, the ejection criterion was $100$~AU from the primary star.

\begin{figure*}[htp]

  \centering

  \begin{tabular}{cc}


    \includegraphics[width=85mm]{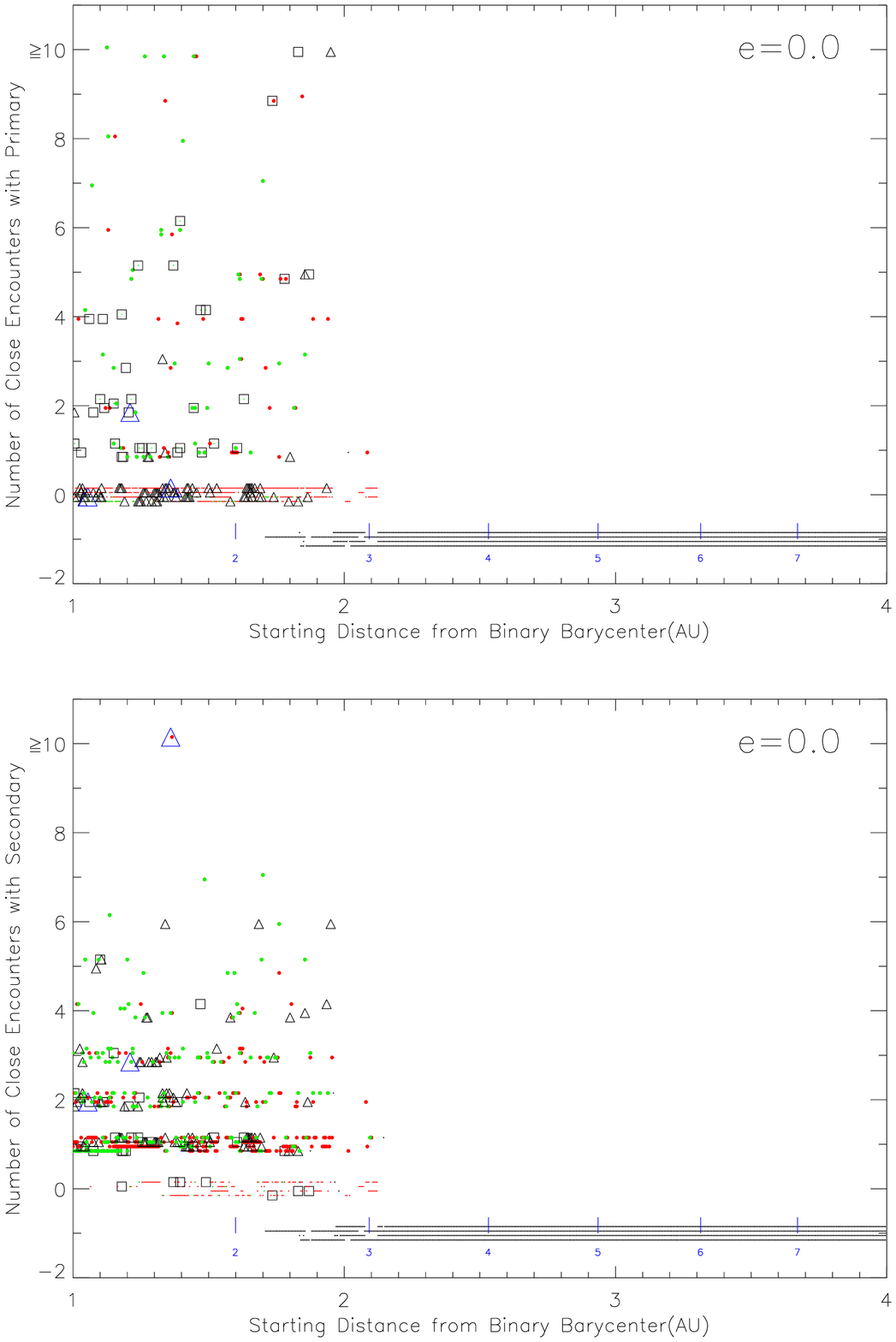}&

    \includegraphics[width=85mm]{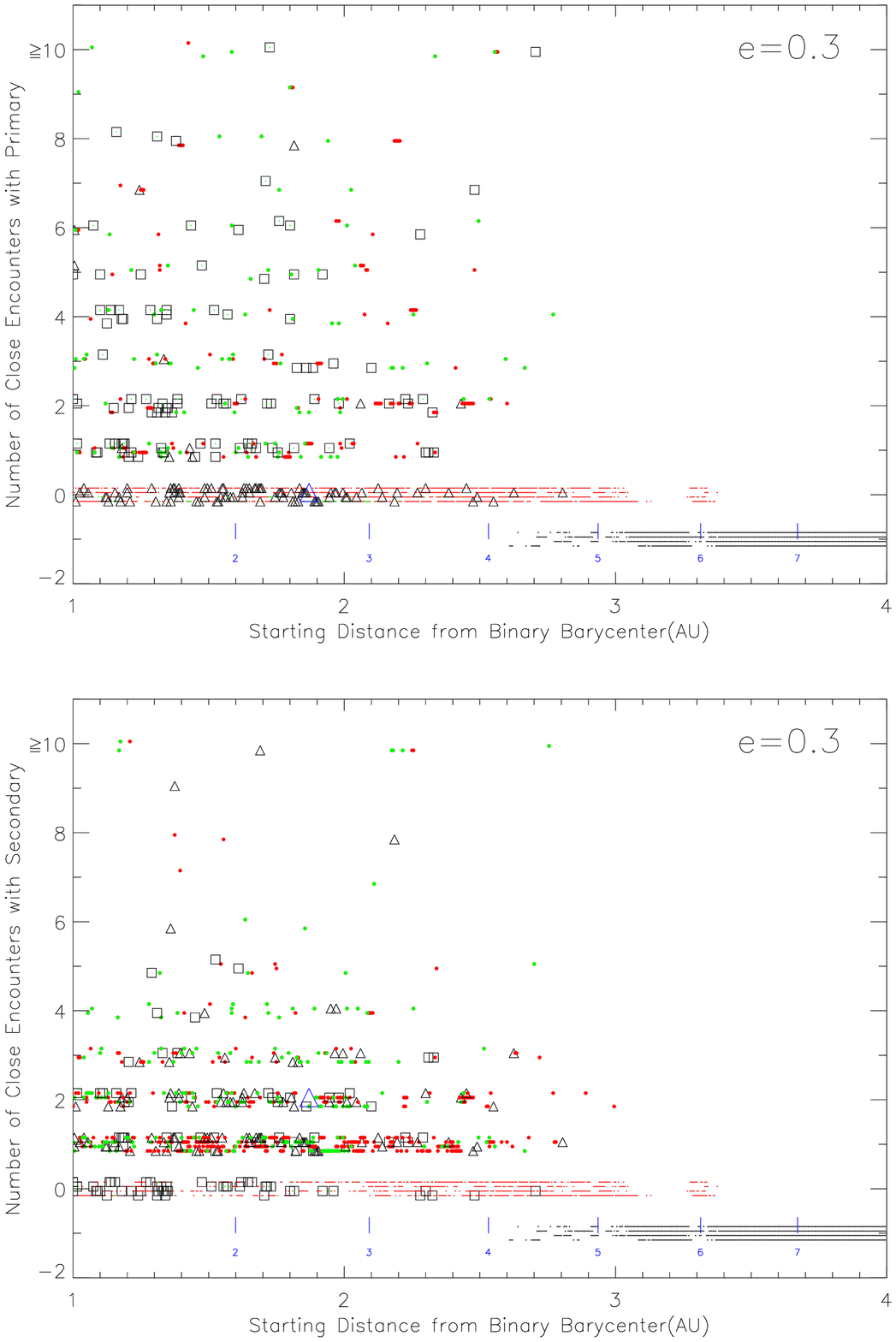}

  \end{tabular}

\caption{Number of close encounters with the primary {\it top} and secondary {\it bottom} versus initial distance from binary, similar to figure~\ref{fig:e0}, but this time allowing for capture by tidal stripping.  {\it Left:} binary $e=0.0$; {\it Right:} binary $e=0.3$. Captures are shown with a large blue symbol, square for those captured by the primary and triangle for those captured by the secondary. Planets that underwent tidal interactions with one of the stars and were ejected are marked green. Planets that underwent tidal interactions and impacted one of the stars are marked black. For captures, the number of close encounters is the number of passages below $0.05$~AU prior to obtaining an orbit with a semimajor axis less than $0.05$~AU. \label{fig:e0tide} \vspace{0.36 in} }

\end{figure*}

\subsection{Results}

We ran the same initial conditions grid as section~\ref{sec:traj}, now with the tidal stripping prescription.  The results of the planets' fate were altered by the addition of the tidal force (figure~\ref{fig:e0tide}).  Now, besides ejection and collision, capture is a possibility. The tidal energy change altered the number of close encounters for planets that already had close encounters, and it resulted in some captures for both the primary and secondary. The majority of the planets with close encounters also had a tidal stripping event.  Ejections were also often preceded by stripping events.  Sometimes the first close encounter was less than the physical radius of the star, i.e., an inevitable collision has already occurred, in which case we did not record that the planet was stripped on its way in. A small minority of the changes in the results is due to the adaptive time step in {\it Mercury} and the chaotic nature of the system. 

Adding tides changed the impact rate for the secondary, $12.97 \pm 0.49 \%$ in both the case without tides and $8.45 \pm 0.54 \%$ in the case with tides, but did not alter the primary impact rate significantly, $6.05 \pm 0.35 \%$ in the case without tides and $6.11 \pm 0.36 \%$ in the case with tides.

Of all the unstable planets in the 1 AU wide binary, approximately $0.16 \pm 0.06 \%$ were captured by the secondary. The primary did not capture any unstable planets. Planets that had multiple close encounters were more likely to become captured than those with only a few. For the low eccentricity case, the planets that impacted the primary typically also had a close encounter with the secondary. The inverse is not true: many planets encounter only the secondary. While this is still true for the higher eccentricity case, there are more planets that manage to impact or be captured by the primary without a close encounter with the secondary.

\subsection{Wide Binary Case}

\begin{figure*}[htp]

  \centering

  \begin{tabular}{cc}


    \includegraphics[width=85mm]{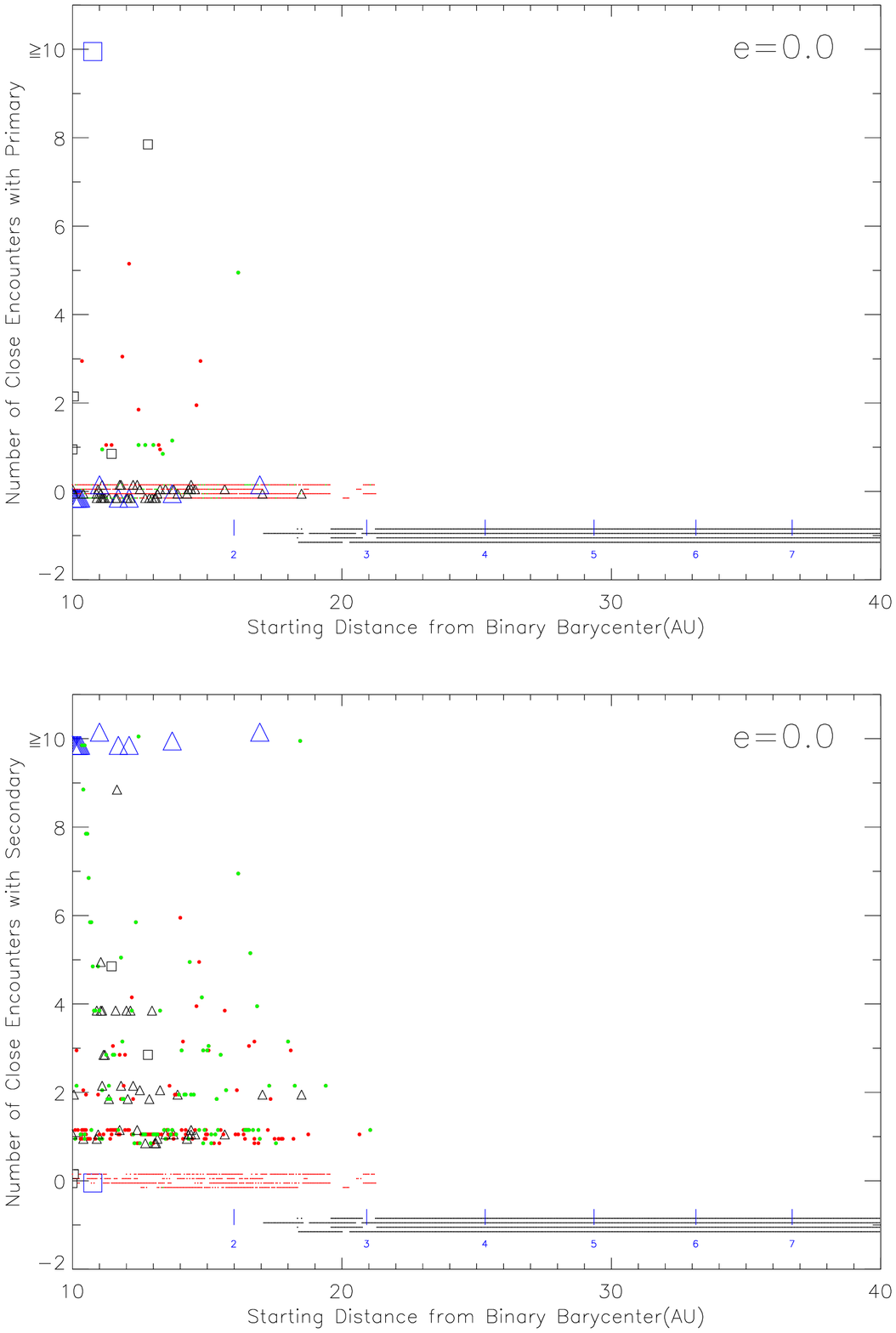}&

    \includegraphics[width=85mm]{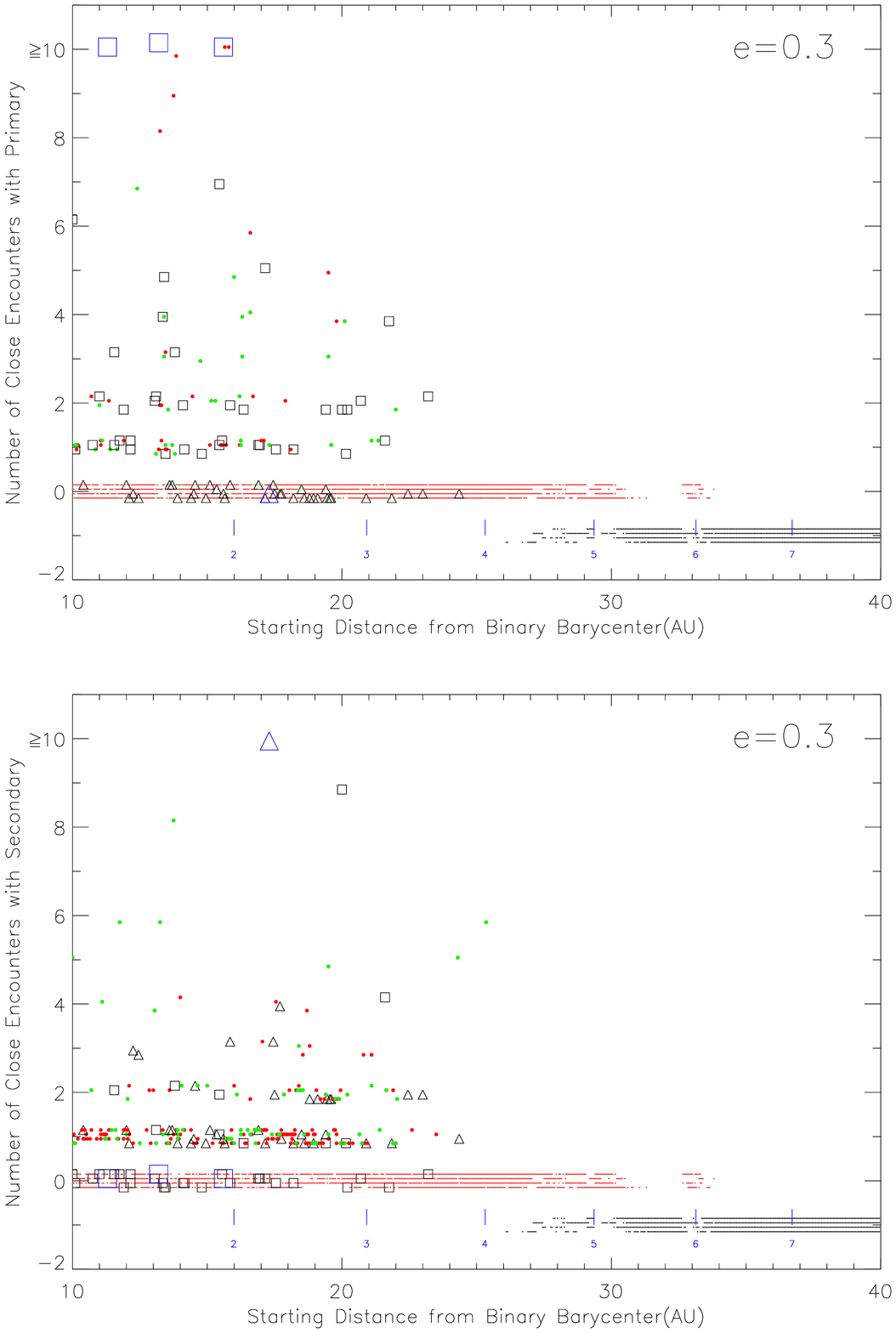}

  \end{tabular}

\caption{Number of close encounters with the primary {\it top} and secondary {\it bottom} versus initial distance from binary, similar to figure~\ref{fig:e0tide}, but with a wider binary.  {\it Left:} binary $e=0.0$; {\it Right:} binary $e=0.3$. Captures are shown with a large blue symbol, square for those captured by the primary and triangle for those captured by the secondary. Planets that underwent tidal interactions with one of the stars and were ejected are marked green. Planets that underwent tidal interactions and impacted one of the stars are marked black. For captures, the number of close encounters is the number of passages below $0.05$~AU prior to obtaining an orbit with a semimajor axis less than $0.05$~AU. \label{fig:abin10} \vspace{0.4 in} }

\end{figure*}

We repeated the simulations for another case of a wider binary separation of 10 AU, see figure \ref{fig:abin10}. The setup was identical in method to the previous runs but the simulations ran for about $3.2 \times 10^5$ years, an equal number of binary periods to the close binary case, and the planets' starting positions ranged from 10 AU to 40 AU with spacing of 0.05 AU and the central body was placed at 50 AU from the barycenter. 

The wider binary produced a similar number of unstable planets, only an increase of $5.6 \%$, compared to the closer binary with tides. A notable difference is the appearance of captures around the primary in the wider binary, although only $0.15 \pm 0.06 \%$ of unstable planets find a stable orbit around the primary. The wider binary produced a significant increase in captures around the secondary as well, up from $0.16 \pm 0.06 \%$ in the close case to $0.48 \pm 0.10 \%$. The wider binary also lowered the number of collisions with the stars from $6.11 \pm 0.36 \%$ for the primary and $8.45 \pm 0.42 \%$ for the secondary to $2.30 \pm 0.22 \%$ and $3.56 \pm 0.27 \%$. The ejection rate changed from $85.28 \pm 0.42 \%$ in the close case to $93.52 \pm 0.36 \%$ in the wide case. 

\epsscale{1.2}
\begin{figure}
\plotone{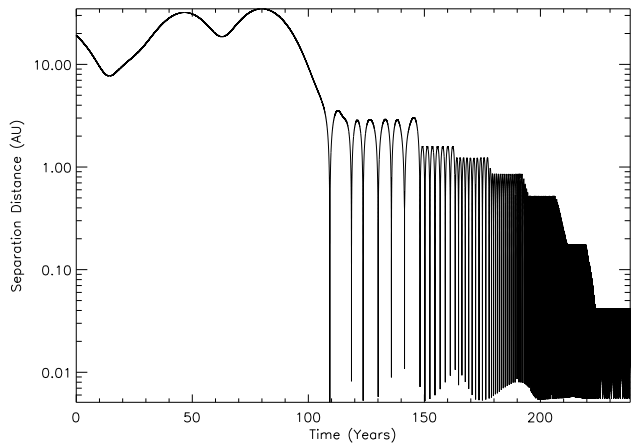} 
\caption{  
Distance from secondary versus time. A typical capture of a planet starting 16.85 AU from the barycenter around the secondary in the 10 AU wide binary.
\vspace{0.4 in} }
\label{fig:capture}
\end{figure}
\epsscale{1.0}

By decreasing the relative size of the stars radius in relation to it's orbit, in the case of the wide binary, collisions become less frequent. A typical capture event is depicted in figure \ref{fig:capture}. Captures become more common since the tidal interactions are based on a fixed amount of energy corresponding to a characteristic tidal distance and in this wider binary, the energy change due to tides is larger compared to the energy of the orbit. The total orbital energy of the close binary is $9.8*10^{43}$, while the energy of the wide binary is $9.8*10^{44}$. The maximum energy a tidal encounter can change a planet's orbital energy by is $4.381×10^{42}$ erg for tides as max. The ratio of maximum orbital energy change to binary orbital energy is $4.5$ \% for the wide binary but only $0.45$ \% for the close binary.

\vspace{.5cm}
\section{Discussion} \label{sec:disc}

We have tracked the fates of circumbinary planets via numerical simulations.  Aside from ejection, we found a substantial number of collisions on the host star.  Jupiter-type planets may pass near enough to their stars that their envelopes are tidally stripped.  This stripping, however, does not appear to alter the orbit enough to capture the planet. 

The fate of circumbinary planets is ultimately dependent on planet formation. Our simulations assume an evenly distributed number of planets in locations where planet formation around circumbinary is unlikely. There exist certain regions where planet formation is unlikely. \cite{Meschiari2012} investigated planet formation around Kepler-16 b, a relatively co-planar system similar to that of our study, and suggest Kepler-16 b formed close to its current position or formed significantly outside and migrated inward and more recent investigations, \cite{2014Lines} and \cite{2012Paarde}, suggest that the majority of discovered circumbinary migrated in to their current locations. If the majority of circumbinary planets form outside and migrate in to their final resting place, the fate of the unstable circumbinary planets would be similar to that of the most distant unstable planet. The most distant unstable planets are located in the island of instability in integer resonances, all of which are ejected without any close encounters with either star, suggesting that  planets migrating inward might become unstable and be ejected. Depending on the rate of migration, circumbinary planets might be able to move through the island of instability and enter a region where unstable planets are more likely to have close encounters with the binary, increasing the chance of impact and the remote possibility of capture. Figure \ref{fig:lifetime} shows the time for ejection for planets in the islands of instability is between $10^3$ and $10^4$ years for eccentricity 0.1 and above. Suggesting that planets that migrate through the turbulent region faster than $10^3$ years can survive and potentially impact or become captured around a single star or exist in a stable orbit inside of the island of instability. \cite{2013Popova} suggests planets can move through the chaotic integer resonances since a number of circumbinary planets thought to have formed at some distance and migrated in have been found in between two chaotic resonances.

Just after formation of planets around binaries, there may be a substantial gaseous disk that can both cause migration of the planet and also feed disks around each of the stars.  In principle, these circum-primary disks could be another source of dissipation to the planet's orbital energy.  In particular, \cite{2009FontRibera} discussed how highly eccentric and distant planets might be captured by circumstellar disks to draw them closer to the star.  A similar process might capture planets into orbit around the individual stars in binaries.

\subsection{Free Floating Planets}
\label{sec:ffp}

Unstable circumbinary planets may be a source of recently discovered free floating planets \citep{2011Sumi}. This mechanism is analogous to the planet-planet scattering mechanism proposed by \cite{2009Veras}. A majority of unstable circumbinary planets were ejected, many of which experienced a tidal stripping event. Depending on the method of planet formation and migration, the rate of ejected planets might be even larger since the rate of ejection is much higher than collision for the planets with the greatest distance from the binary. Migrating planets will be ejected without any tidal stripping events if the rate of migration is slower the the time of ejection, figure \ref{fig:lifetime}, in the island of instability, where almost all planets were ejected without any tidal encounters. A possible method to determine if free floating planets were once circumbinary would be detect the evidence of a tidal encounter. A potential signature would be the fast rotation rate expected for the remnant of such an interaction \citep{2005Faber}. On the contrary, if a candidate free-floating planet is found to have satellites or a binary companion, then it has not endured a tidal stripping event.

\vspace{.3 in}

\subsection{Stellar Pollution}

The history of both collision and tidal events should be spectroscopically visible through the pollution of planetary material in either star. Close passes that experience tidal stripping events should pollute the star provided there is enough gas present to be stripped. Collision can also be visible by determining the abundance of elements due to the addition of planetary material. By comparing the amount of planetary material in both stars in the binary, it would be possible to confirm the expected rates that unstable circumbinary planets impact the primary and secondary.

It is possible to determine the iron abundance differences for wide binaries and from which perhaps determine a preference of collision for unstable stars. Studies like \cite{2004Desidera} can rule out the pollution of mass as small as $1$ Earth Mass of iron. The accuracy of the study of iron abundance differences in binaries is constrained by the temperature of the binaries and the size of the convective zone of the star. According to \cite{2006Desidera}, the difference in chemical abundances in the wide binaries they studied are not due to the ingestion of planetary material. Similar studies of close binaries do not currently exist and might prove to be impossible to resolve spectroscopically.

\acknowledgements  We thank Kaitlin Kratter and Rosemary Mardling for helpful conversations and insightful comments. 

\bibliography{mscbp.bbl} \bibliographystyle{apj}

\end{document}